# Population Density and Spreading of COVID-19 in England and Wales.

Jack Sutton[1], Golnaz Shahtahmassebi[1], Haroldo V. Ribeiro[2], and Quentin S. Hanley[1*]

[1] *School of Science and Technology,*
*Nottingham Trent University,*
*Clifton Lane,*
*Nottingham,*
*United Kingdom*

[2] *Departamento de Física,*
*Universidade Estadual de Maringá,*
*Maringá,*
*Brazil*

[*] Corresponding author: Q. S. Hanley

E-mail: quentin.hanley@ntu.ac.uk

# Abstract

We investigated daily COVID-19 cases and deaths in the 337 lower tier local authority regions in England and Wales to better understand how the disease propagated over a 15-month period. Population density scaling models revealed residual variance and skewness to be sensitive indicators of the dynamics of propagation. Lockdowns and schools reopening triggered increased variance indicative of outbreaks with local impact and country scale heterogeneity. University reopening and December holidays triggered reduced variance indicative of country scale homogenisation which reached a minimum in mid-January 2021. Homogeneous propagation was associated with better correspondence with normally distributed residuals while heterogeneous propagation was more consistent with skewed models. Skewness varied from strongly negative to strongly positive revealing an unappreciated feature of community propagation. Hot spots and super-spreading events are well understood descriptors of regional disease dynamics that would be expected to be associated with positively skewed distributions. Positively skewed behaviour was observed; however, negative skewness indicative of "cold-spots" and "super-isolation" dominated for approximately 8 months during the period of study. In contrast, death metrics showed near constant behaviour in scaling, variance, and skewness metrics over the full period with rural regions preferentially affected, an observation consistent with regional age demographics in England and Wales. Regional positions relative to density scaling laws were remarkably persistent after the first 5-9 days of the available data set. The determinants of this persistent behaviour probably precede the pandemic and remain unchanged.

# Introduction

SARS-CoV-2 spread rapidly from a cluster of cases in China in late 2019 to a global pandemic on 13 March 2020. The number of confirmed cases of COVID-19 continues to grow worldwide with over 186 million cases and over 4 million deaths. SARS-CoV-2 is thought to spread by direct contact, fomites, and aerosols from both symptomatic and asymptomatic people [1–4]. During the pandemic, distancing measures and meeting size restrictions have been widely deployed to slow the spread of the disease by reducing the number and duration of interactions capable of causing infection. At scale, population density could be a proxy for these interactions. For example, someone living in a region of high population density is expected to have a greater number of interactions compared with someone who lives in a rural setting [5].

The propagation of COVID-19 via super-spreading events has been documented [6–10] these events are reported to have fat tails and distributions presented show strong positive skew. For subsequent modelling a range of distributions have been used including Weibull [11], Poisson [12], gamma [7], and normal [13] and subsequent modelling is primarily based on population. The effects of population size on COVID-19 dynamics have been investigated previously including aspects of population density effects [14–17]. Investigations of population density effects have been limited to a relatively small number of time points aggregated over a period of time, usually a month or year [18–24]. Daily granularity of data is not easily accessible; however, the COVID-19 pandemic has provided a unique and evolving data set with daily updates for generating an extended scaling time series. These data have been influential in informing government interventions, policy decisions, and public perceptions allowing data driven informed decisions [25].

These daily data at relatively high regional granularity provide an opportunity to document the daily evolution of scaling metrics, descriptive statistics, and residual variance over an extended period. Here, we investigated scaling behaviour in England and Wales using daily COVID-19 cases and deaths in England and Wales Lower Tier Local Authorities (LTLAs) with population density. These were examined to better understand how infectious disease metrics progress over time at country scale.

## Scaling Models

Urban scaling [26] considers population to predict a range of urban indicators. A variety of mathematical forms have been applied with power laws being widely used.

$$Y = Y_0 P^\beta \varepsilon \quad (1)$$

Here, $Y$ is the indicator, $P$ is the population, $\beta$ is the scaling exponent, $Y_0$ is the pre-exponential factor and $\varepsilon$ are residuals that are independent and identically distributed with common $N(0, \sigma^2)$ distribution. An estimate to the parameter $\beta$, can be obtained by applying the least square method to the logarithmic version of equation 1 (*i.e. logY vs. logP*) which aims to minimise the value $\sum \hat{\varepsilon}_i^2$.

When combining rural and urban regions, density metrics provide better models [20–21] than population. This can be described by similar power-law functions of the form

$$Y_D = Y_0 P_D^{\beta_D} \varepsilon \quad (2)$$

where $Y_D$ is the indicator density, $P_D$ is the population density and $\beta_D$ is the density scaling exponent. Indicator and population densities are obtained by dividing them by the corresponding defined regional area, $A$ (*i.e.* $Y_D = Y/A$ and $P_D = P/A$). Similarly, to population scaling, when $\beta_D < 1$ scaling is sub-linear, when $\beta_D = 1$, the scaling is linear and when $\beta_D > 1$ the scaling is super-linear. When interpreting density scaling results, sub-linear scaling accelerates in rural (low-density) regions and super-linear scaling accelerates in urban (high density) areas. The log transformed data is usually fitted to the logarithmic form

$$\log(Y_D) = \log Y_0 + \beta_D \log(P_D) + \varepsilon \quad (3)$$

to obtain the regression model parameters.

Residuals, $\varepsilon_i$, from the fit to the model defined in equation (3) are obtained using least squares method which aims to minimise the variance $\sum \varepsilon_i^2$ for

$$\varepsilon_i = \log(Y_{D,i}) - \log(\hat{Y}_{D,i}) \quad (4)$$

for $i=1, ..., n$ and $\log(\hat{y}_{D,i})$ is the estimate of $\log(Y_{D,i})$. Negative values of $\varepsilon_i$ are below expectation while positive $\varepsilon_i$ are above expectation.

## Residual and Case Density Models

The distribution of residuals obtained from the England was modelled using normal and generalised logistic (GL) distributions. The latter has the form,

$$GL(x; \theta, \sigma, \alpha) = \frac{\alpha}{\sigma} \frac{e^{-\frac{x-\theta}{\sigma}}}{\left\{1 + e^{-\frac{x-\theta}{\sigma}}\right\}^{\alpha+1}} \quad (5)$$

where $\vartheta$, $\sigma$ and $\alpha$ are the location, scale and shape parameters respectively such that $\alpha>0$, $\sigma>0$ and $-\infty<x<+\infty$. The first moment of the GL is $E(X) = \vartheta + \sigma(\Psi(\alpha) - \Psi(1))$ where $\Psi(1) \cong -0.57721$. The second moment of the GL distribution is $Var(X) = \sigma^2(\pi^2/6 + \Psi'(\alpha))$. The GL distribution was selected due to its flexibility modelling data with a range of different shapes under a single framework.

## Materials and Methods

### Data Sets

English and Welsh data on the number of daily COVID-19 cases and English deaths were obtained from Public Health England (PHE) (https://coronavirus.data.gov.uk/) for lower tier local authorities (LTLAs). Wales have a different methodological approach in their death data collection and therefore excluded in any of the death analysis within this study. Meanwhile English death statistics in this study are people who had a positive test result for COVID-19 and die within 28 days. COVID data are available is a range of time and spatial scales from both PHE and the UK Office of National Statistics (ONS). Data from PHE was available at middle super output area (7,210 regions) but on a weekly basis. ONS does surveys of prevalence, however these also do not have daily granularity. We selected the daily data as the best compromise between temporal and spatial coverage as well as allowing the most up-to-date coverage. England and Wales population estimates were based on the 2011 census and regional land areas were obtained from NOMIS (https://www.nomisweb.co.uk), a database service run by the University of Durham on behalf of the UK Office for National Statistics. The shape files for LTLAs were obtained from the open geography portal (http://geoportal.statistics.gov.uk) provided by the UK Office for National Statistics and UK Data Service (https://census.ukdataservice.ac.uk). LTLAs for COVID-19 cases (in England and Wales), COVID-19 mortality (England alone), population, and area were aligned in a daily time series covering the period from 01/03/2020 to 20/05/2021. All data in this study are publicly available under Crown Copyright. The data are provided by PHE such that counts between 0 and 2 are blank in all CSV files and NULL in any other formats. These adaptations in the data are to prevent disclosure issues. In the analysis, zero values representing null returns were considered as missing values. The population in LTLA data for City of London and Isles of Scilly are considered small and therefore statistics in these regions were combined with Hackney and Cornwall, respectively. The data are limited by the conditions in place at the time they were reported. Specifically, the availability of tests was limited in the earliest period and changed greatly over the period. The limitations created by PHE disclosure control, null data, combined regions and variations in testing are inherent in the data set.

## Statistical Analysis

The data were analysed using the statistical software R version (3.6.2) [27] with the sf (0.9-1) [28], raster (3.0-12) [29], dplyr (0.8.5) [30], spData (0.3.5) [31], tmap (2.3-2) [32], ggplot2 (3.3.0) [33–36], xlsx (0.5.7) [37], gplots (3.0.4) [38], httr (1.4.2) [39], plyr (1.8.5) [40], png (0.1-7) [41], rgdal (1.5-19) [42], rgeos (0.5-5) [43], lubridate (1.7.9.2) [44], fitdistrplus (1.1-3) [45], fgarch (3042.83.2) [46] and glogis (1.0-1) [47] packages.

# Results and Discussion

## Overview of regions, cases, and number of observations

England and Wales have 337 LTLAs (315 English LTLAs and 22 Welsch LTLAs) which range in area from 1213 ha (Kensington and Chelsea) up to 518,037 ha (Powys) and have populations between 37,340 (Rutland) up to 1,070,912 (Birmingham). Population densities vary from 0.25 people per hectare (p/ha) (Eden) to 138 p/ha (Islington). Not all LTLAs reported cases or deaths on each day within the period leading to variability in observations (Fig 1). This largely tracked the general progress of the pandemic with the summer months showing the fewest cases, deaths and observations. Histograms of per capita cases (Fig 2) exhibited variable shapes over the course of the pandemic with some periods showing negative skew (Fig 2(a)) while at others they were positively skewed (Fig 2(b)). The availability of testing varied widely over the 15 months which may be a confounder in some presentations; however, the daily scaling metrics, variance, and skewness will reflect the processes in place on the day and were not obviously aligned with testing or the number of observations. All daily per capita case histograms can be found in Fig S1 in the supplementary material.

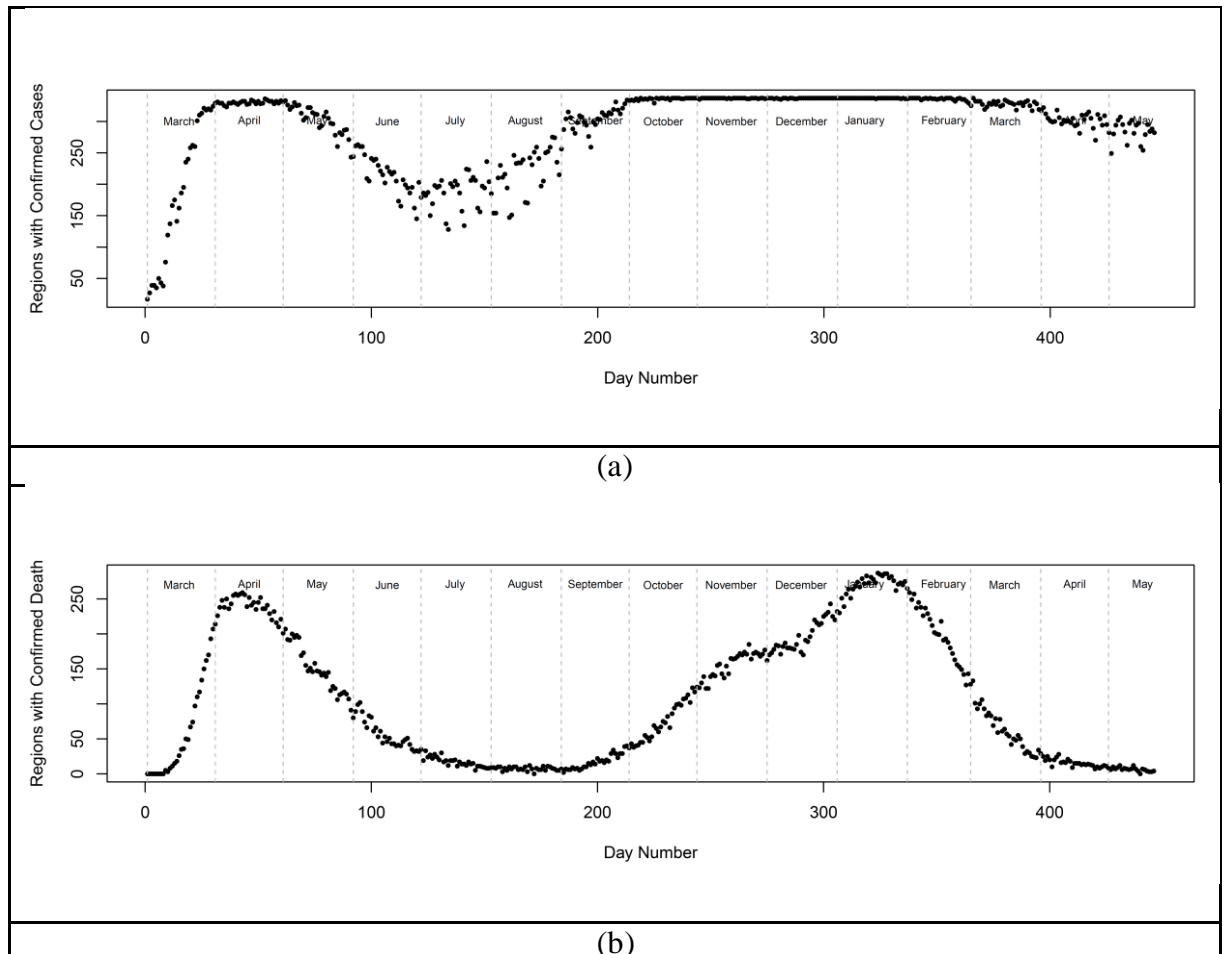

(a)

(b)

**Fig 1. Time series indicating the number of LTLAs returning cases (a) or deaths (b) over the period of study.**

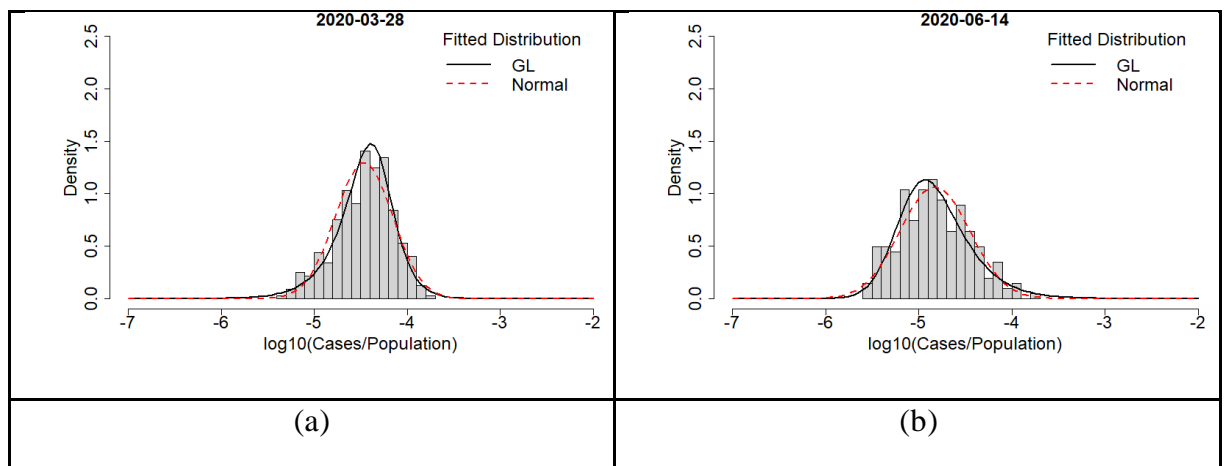

**Fig 2. Histograms of per capita cases in English and Welsh LTLAs.** Some periods within the time series showed negative skew (a) while others were positively skewed (b).

## Daily Progression of COVID-19

To test for scaling behaviour and to correct for the known bias of *per capita* measures daily scaling plots (Figs 3 (a-d): and Figs S2 and S3) were constructed and found to be consistent with single power-law models throughout the pandemic. The daily residuals obtained were used as scale adjusted metrics to create geomaps (Figs 3 (e-h): Figs S4 and S5). Residuals are more useful metrics that could be used to assist local interventions.

In the scaling plots (Figs. 3 (a-d)), variability in residual variance was clear by inspection. For example, toward the end of the December holiday period (25/12/2020; Fig 3(c)) the data were closer to the power law than in September (16/9/2020; Fig 3(b)). The low variance periods represent a more homogenous presentation of cases across the regions while the higher variance periods were indicative of more heterogeneous regional cases. All daily scaling plots and corresponding geomaps can be found in S2-5 Figs provided in the supplementary material.

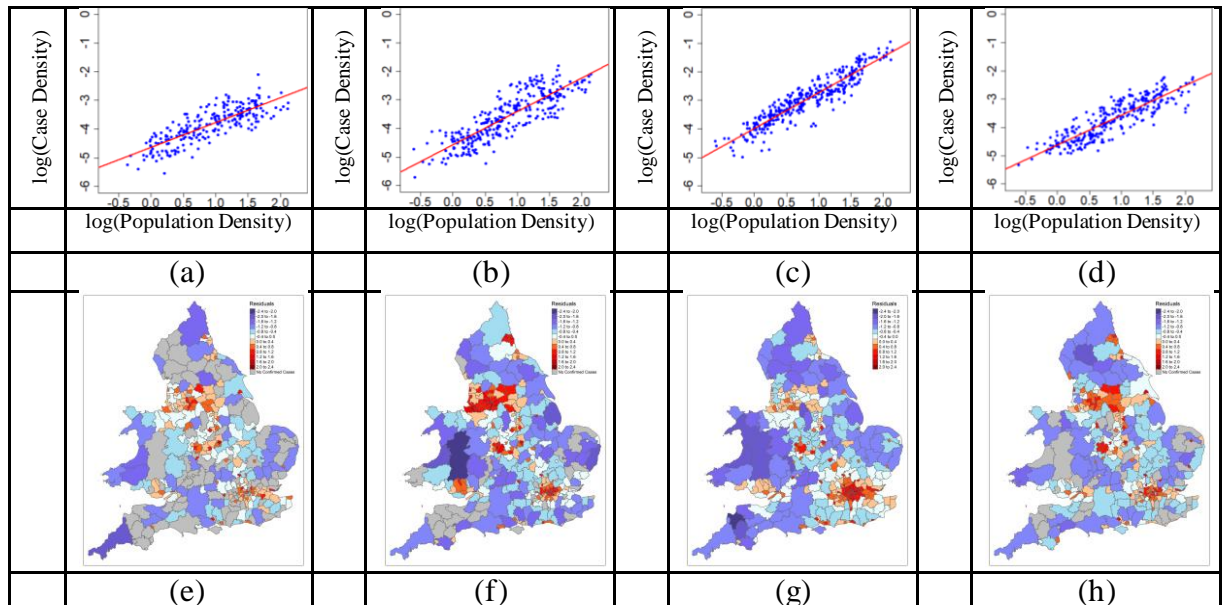

**Fig 3. Scaling plots and geoplots at different times during the pandemic.** These are recorded on the (a and e) 08/06/2020. (b and f) 16/09/2020. (c and g) 25/12/2020. (d and h) 04/04/2021. Regions that are red are above expectation and blue is below. The darker the shade the further from the scaling law.

## Daily Exponent, Variance, and Skewness for Cases

The LTLA data were examined to assess the trajectory of scaling behaviour, residual variance, and skewness over 15 months of the pandemic for cases (Fig 4). The scaling exponents (Fig 4(a)) for cases rose quickly reaching a peak near the beginning of the first lockdown (announced on the 23/03/2020) in the England and Wales and declined gradually until restrictions were slowly eased toward the end of May and early June. The peak in cases during the first three months coincides with super-linear scaling; however, super-linear scaling was not universal and the preference for propagation in rural vs. urban regions reversed three times during the period of study: early-March, late April, and the end of July. Although periods of rapid growth, seem to coincide with acceleration in high population density regions the long-term behaviour of the pandemic makes clear there is no universal rule and population density is not a simple proxy for infectious interactions.

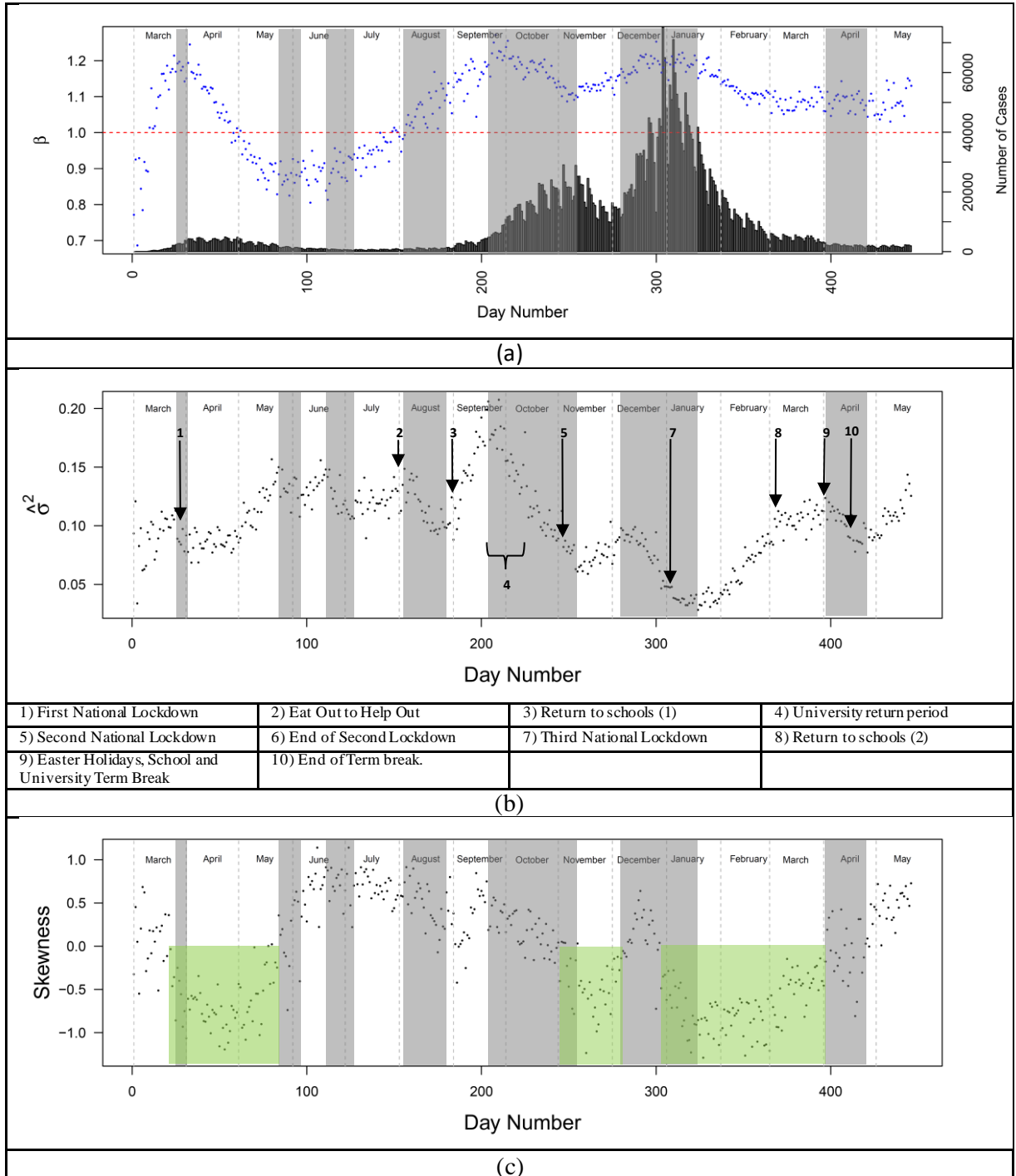

**Fig 4. Daily time series of scaling exponent and residual variance and skewness for cases between 01/03/2020 and 11/01/2021.** (a) Time series of daily scaling exponent of COVID-19 cases, (b) residual variance, and (c) residual skewness. The horizontal line in (a) indicates linear scaling. The bar chart indicates raw daily cases. The grey shading indicates periods of homogenisation. The green shaded periods in (c) indicate times when isolation dominates (negative skewed residuals). The remaining times were dominated by spreading (positively skewed residuals). Arrows indicate key dates/time periods. The national restrictions in Wales preceded England beginning on 20/10/2020.

The trajectory of residual variance (Fig 4(b)) changed by over a factor of 4 during the 15 month period and presented a contrast to the scaling parameters. Variance remained relatively constant until late April and the later stages of the first lockdown when regional heterogeneity increased distances from the scaling law. Released restrictions tended to homogenize and reduce the distance of individual regions to the scaling law. In the summer months, regional lockdowns (Leicester and greater Manchester) in late June were followed by increasing variance. Residual variance grew rapidly following the opening of schools around (First week of September), doubling in approximately 2 weeks suggesting that schools assisted in propagating cases, but the impact was mostly contained within their immediate regions. Although an immediate "surge" in cases was not seen, a continuation of a consistent increase in cases that began in August was observed. The increasing variance indicates heterogeneous propagation that continued until mid-September when the trend reversed until the beginning of the second national lockdown in November (05/11/2020).

The period of declining variance and homogenisation coincides with students returning to universities. There are approximately 2.4 million students attending universities in the UK. University teaching terms have staggered start dates from the last weeks of September through the first weeks in October. These typically follow a week of orientation and social activities. In advance of orientation and the start of teaching, many students travel with their families from all parts of the UK along with a large number of students who arrive from abroad. This process changed the dynamics of propagation in England and Wales during this time. While there may be other explanations in addition to universities re-opening, there are no other obvious country scale policy changes or processes during this time window.

Homogenisation also occurred following the release of the national restrictions (03/12/2020) and the reopening of businesses in the second national lockdown (12/04/2021). Notably, only the abrupt release of the national restrictions is associated with an obvious "surge" in cases.

This includes the major holidays of Christmas and New Year's. Neither caused a "surge." They continued the propagation of the disease in a way that was consistent before and after these key dates. The general country scale homogenisation between the LTLA regions drove residual variance to the lowest levels seen over the 15-month period.

Skewness provides a further contrast to case counts, scaling exponents, and variance. We used the scaling law residuals to create a time series of skewness metrics (Fig 4(c)). Similar behaviour was seen in the *per capita* case distributions (Fig 2) with characteristics changing over the course of the 15-month period. When cases follow a distribution with a strong positive skew, the long positive tail of the skewed distribution is indicative of propagation with hot spots and potential super-spreading incidents. Conversely, when the residuals are negatively skewed, this indicates a distribution better characterised by a long tail of "cold spots" or super-isolated regions.

## Daily Exponent, Variance, and Skewness for Deaths

In contrast, daily exponents, variance, and skewness for COVID-19 deaths (Fig 5) were consistent and remained at a similar level throughout the pandemic. For a short time at the beginning of the time series, regions exhibited super-linear scaling, but this inverted circa 10/04/2020 and remained sub-linear thereafter (with the exception of a few days in the summer, and towards the end of the time series when the number of regions with reported deaths are low). This meant that for most of the pandemic rural regions were preferentially affected by deaths. Variance and skewness also do not change much and have very little structure in comparison to cases. This shows constant homogeneous behaviour and similarly distributed deaths throughout England. This is in stark contrast with the far greater structure in cases. This behaviour is consistent with the age demographics in England and Wales. Previous work has documented that populations dense regions serve and a magnet for young people while rural regions tend to have a greater proportion of elderly people [19]. The scaling exponents for deaths throughout are consistent with those seen for scaling of people 60 and above in England and Wales. This is overwhelmingly the demographic most likely to die from COVID-19.

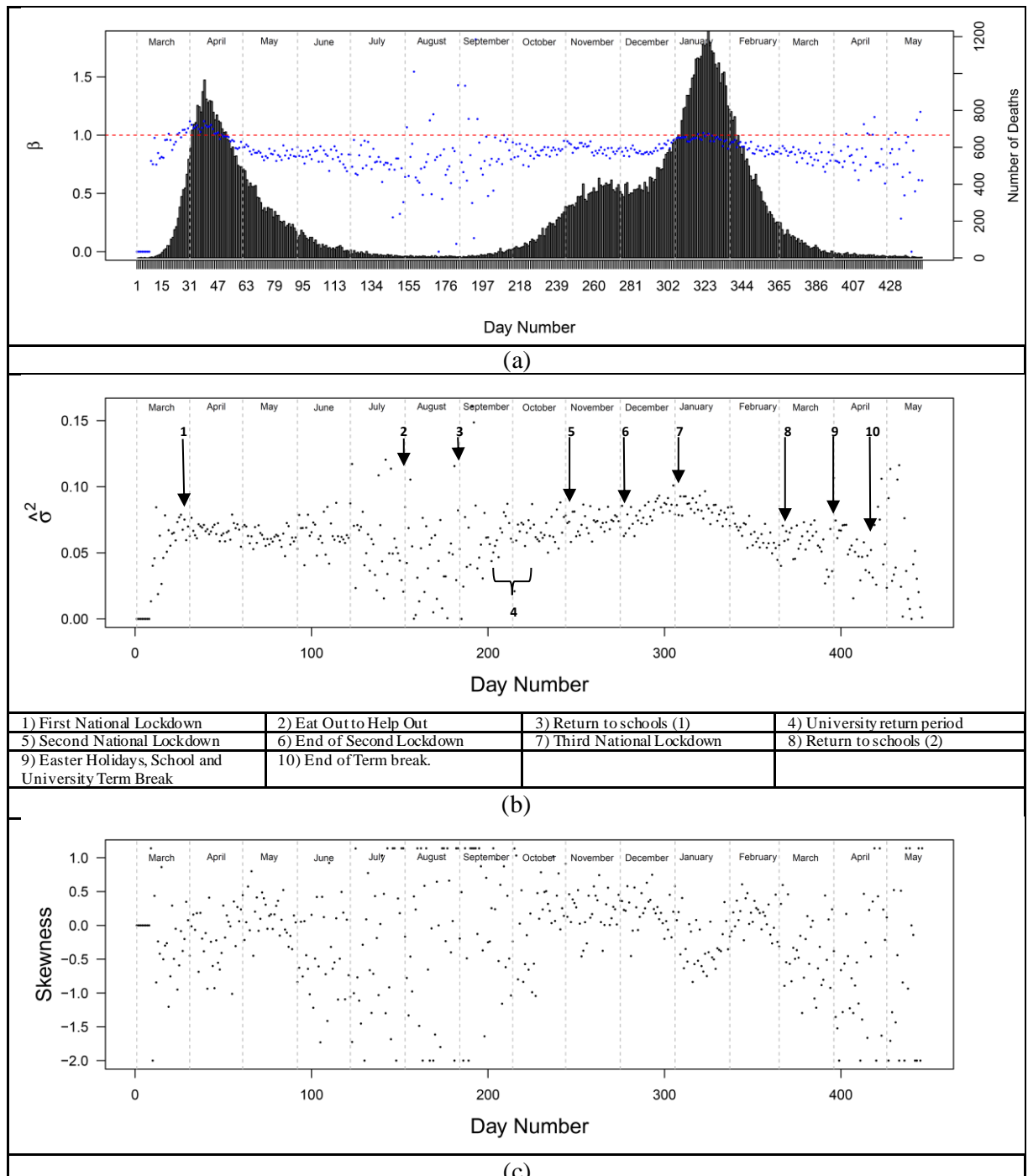

**Fig 5. Daily time series of scaling exponent and residual variance and skewness for deaths 01/03/2020 and 11/01/2021.** (a) Time series of daily scaling exponent of COVID-19 deaths, (b) residual variance, and (c) residual skewness. The horizontal line in (a) indicates linear scaling. The bar chart indicates raw daily deaths. Arrows in (b) indicate key dates/time periods. The second lockdown in Wales preceded England beginning on 20/10/2020.

## Dispersion of COVID-19 Case Residuals over Time

To better understand the distribution of residuals, we investigate the normal and generalized logistic distributions as candidate distributions using the LTLA data (Equation 5). The normal distribution is symmetric and has no skew. The GL distribution has three parameters which can accommodate a wider range of shapes including positive and negative skewing. When comparing normal and GL distributions as models for scaling law residuals the additional parameter needs to be accounted for. We used the Akaike (AIC) and Bayesian (BIC) information criteria to decide if normal or GL represented a better model for each day in the 10-month period. When selecting a model, lower AIC and BIC scores represent better fits. The differences between AIC and BIC scores obtained from fitting the two distributions to the residuals were for each day in the 15-month period (Fig 6). Positive values correspond to a generalised logistic distribution as the preferred model, whilst a negative value corresponds to a normal distribution as the preferred model. All daily histograms for cases and deaths can be found in Figs S7 and S8 in the supplementary material.

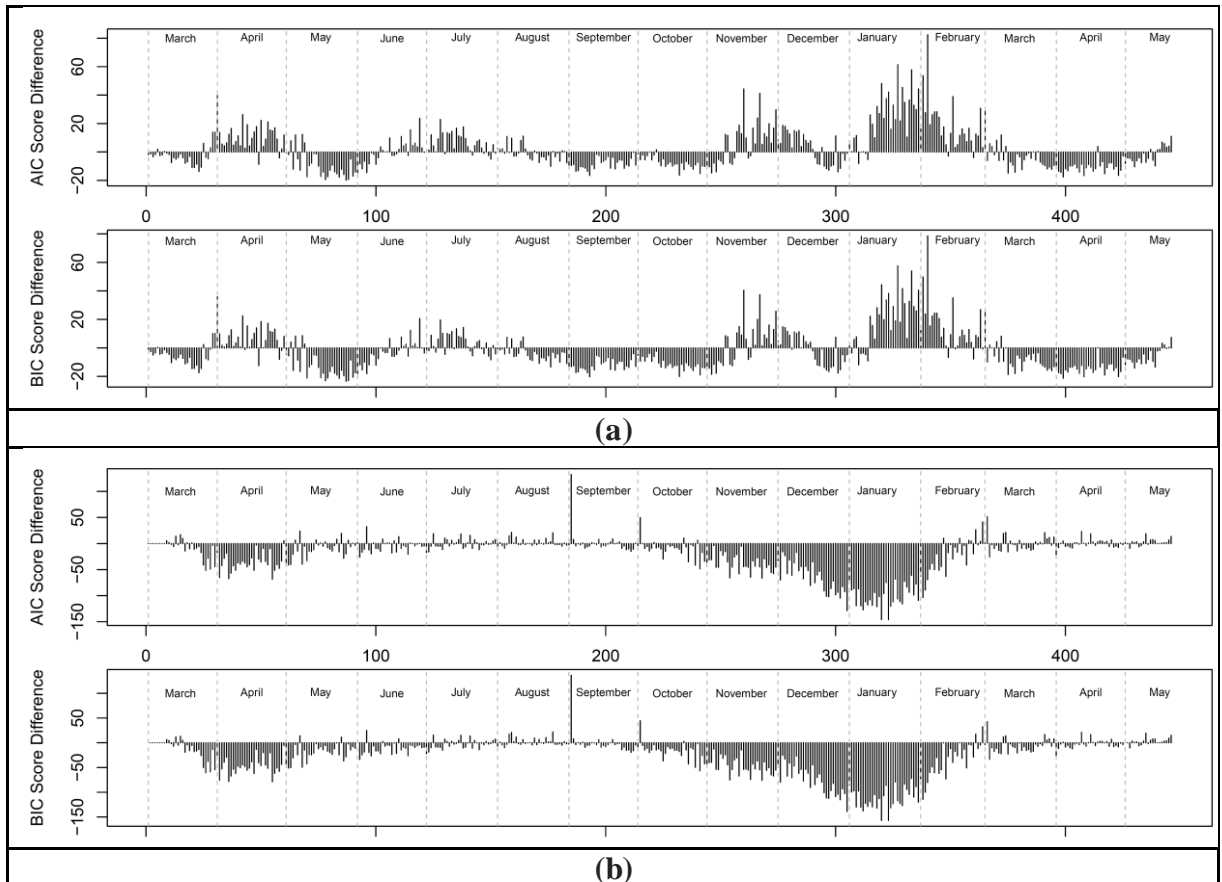

**Fig 6. AIC and BIC differences over time.** (a) COVID-19 cases and (b) COVID-19 death. Positive AIC/BIC indicates GL is a better fit and a negative AIC/BIC indicates normal is a better fit.

Although there is some noise in the differences, the contrast between cases and deaths is again clear. During the initial periods of the lockdowns (March, November and January) propagation of cases was associated with a GL distribution and negative skew whilst during less restrictive time frames (August, September, October and April) propagation is associated with a normal distribution. A variety of authors have noted the fat tails and/or positive skewing in of super-spreading events [6–10]. At LTLA scale, the number and size of individual spreading events place determine its position in a distribution. With a sufficient number of events, it will become a hotspot and appear on the extreme positive side of a positively skewed residual distribution. At times during the pandemic in England and Wales this was observed but was insufficient for the full period. As an example, modelling and simulation of propagation [7] using network science and a gamma distribution was attempted using varying parameter values to represent different proportions of "super-spreaders." This analysis indicated that the initial trajectory of exposed and infected people in a population accelerates quickly in networks where there are a high proportion of super-spreaders. However, a gamma distribution cannot be negatively skewed and the daily cases in the data here contain periods of negative skewing. "Super-spreading" events creating "hot-spots" are certainly important but the converse concepts of "super-isolation" and "cold-spots" better describe regions at the low end of a negatively skewed distribution and this phenomenon needs to be better appreciated and understood. Position relative to expectation was remarkably persistent (Fig S6) and understanding the features of regions where a disease is not spreading or is consistently below expectation is needed. The contrasting behaviour of deaths is also of interest. Residuals more consistent with normally distributed residuals were the overwhelming feature of regional deaths during the 15-month period.

## Regional Persistence of COVID-19 Case Residuals

To investigate the persistence of regional behaviour, the correlation between residuals was computed for all pairs of days and presented as a heatmap (Fig 7). This indicated the position of a particular region relative to the scaling laws was very persistent after the first 5 (Pearson correlation (Fig 7)) to 9 days (Spearman rank correlation (Fig S6)). The near universal dark red appearance of the heatmap indicates that a region that was high relative to the population density scaling law early in the pandemic remained there. While such correlation between close dates is to be expected (Fig 7c) seeing it persist for 300-350 days (Figs 7a and 7b) is remarkable. This persistence which survived 3 national lockdowns, multiple locally targeted measures, and an enormous expansion of testing needs explanation. In previous studies of the inter-relationships between indicators of health, wealth, well-being and age [19], we noted that mortality health outcomes are related to these other factors in complex ways. Based on this earlier work, the socio-economic characteristics leading to a region's position relative to the scaling laws may have been in place before the pandemic began. Further work is needed to test this hypothesis directly.

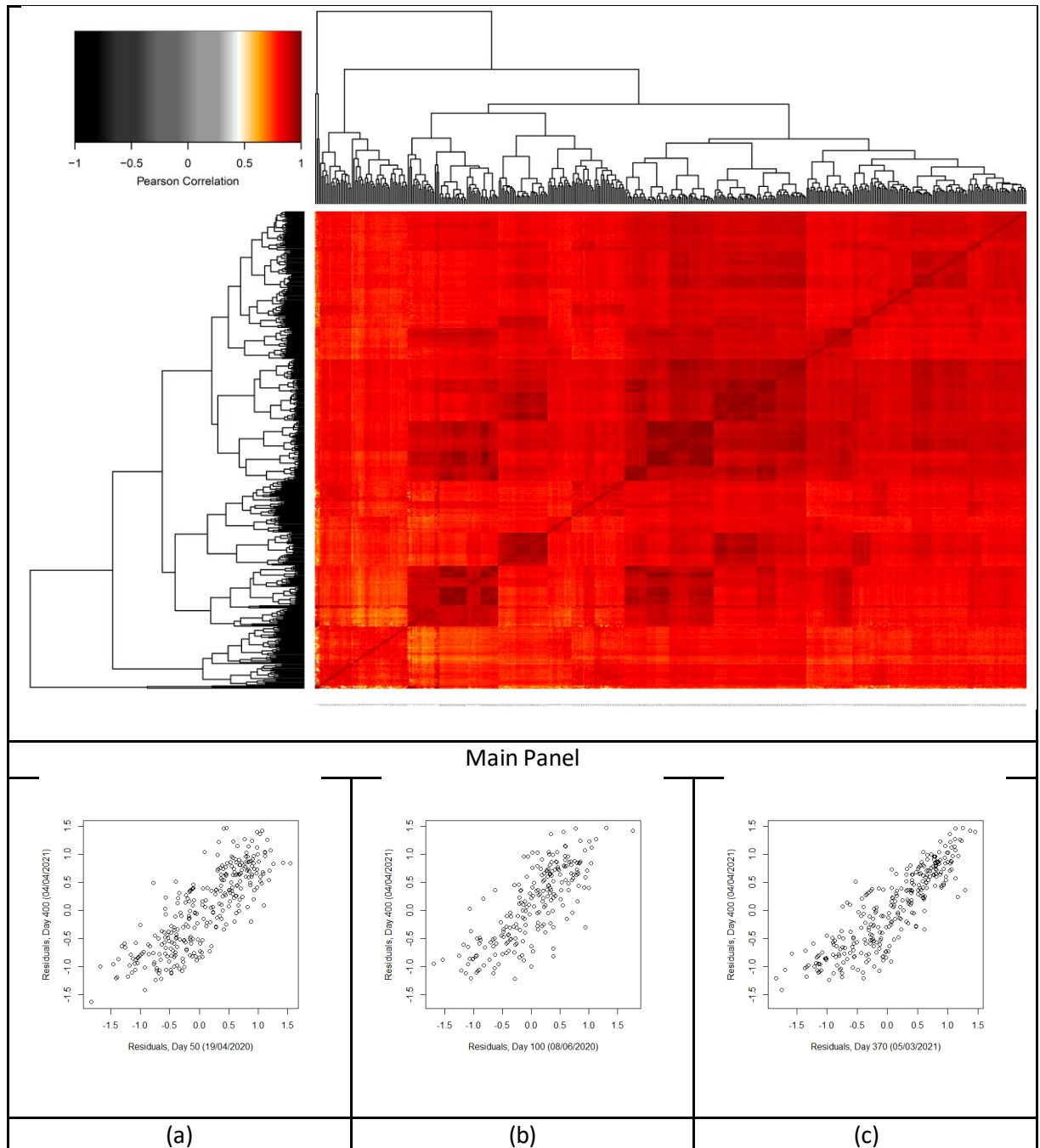

**Fig 7. Heatmap of Pearson's correlation coefficient with example paired dated.** Correlation of COVID-19 case residuals between all pairs of dates between 01/03/2020 and 20/05/2021 (main panel). Red indicates strong positive correlation. White and grayscale indicate low and negative correlation. The darker shade of colour is associated with a higher similarity between the two pairs. Example data sets where residuals from days 50, 100, and 370 are correlated with those from day 400.

# Conclusions

This study has established that both regional *per capita* measures and scaling law residuals exhibit both positive and negative skewing. Positively skewed distributions have been widely observed and used to model pandemic behaviour [7–12]. Such behaviour is important to indicate super-spreading and hot-spots, but insufficient to characterise the full sweep of the pandemic through multiple interventions.

Similarly, scaling law parameters are often thought to be constant or very slowly changing features of a process. In the case of COVID-19 cases, scaling parameters evolved over relatively short periods of time. For cases, the scaling law exponents reached a peak at the beginning of the first lockdown and gradually declined for approximately three months. Preferential propagation of COVID-19 cases switched between rural and urban regions several times during the first 150 days of the pandemic. Since then urban regions have driven cases. COVID-19 mortality gave a more consistent picture of low population density regions preferentially and consistently affected with linear scaling only being approached during two periods in the 15-months studied. Those corresponded with the months of peak death in April 2020 and January 2021.

Variance relative to the population density scaling laws is a key descriptor of the distribution of regional cases. Lockdowns produce heterogeneity (higher variance) across regions while reducing cases. The re-opening of schools drove heterogeneity during a period of case growth indicative of locally important outbreaks. Country scale mixing such as occurred with the opening of universities and holiday periods promotes homogenisation (low variance). All key statistical metrics from regional death data were remarkably different from cases in the time period. This is consistent with regional age demographics in England and Wales. From a policy point of view these observations and patterns are particularly important, as they provide insight and expected indicative effects following implementation of health policies.

Within this framework it is important to note that for the full 15 months period England and Wales had continuous community spread of SARS-COV-2. Excepting the very early period in March, there has been nothing that could be called a "surge." Within the 15-month period, the rise and fall of cases and deaths have been gradual as has the evolution of scaling metrics, variance structures, and distribution shapes.

Finally, regional behaviour relative to population density scaling laws was remarkably persistent. It is possible that the determinants of regional behaviour existed pre-pandemic and although government interventions have had an unambiguous impact on the rise and fall of cases and death, they have had little impact on whether a particular region is high or low in relative to nationwide population density scaling.

# Acknowledgements

The authors are grateful to the UK Office of National Statistics and Public Health England for making these data available.

# Supporting information

**S1 Dataset. LTLA regions included in this study. 337** English and Welsch LTLA's that define regions in this study.

**S2 Dataset. English and Welsch LTLA daily COVID-19 cases employed in this study.** Data covering the period from 01/03/2020 to 20/05/2021

**S3 Dataset. English LTLA daily COVID-19 death employed in this study.** Data covering the period from 01/03/2020 to 20/05/2021

**S4 Dataset. Total number of cases in England and Wales and deaths in England.** Data covering the period from 01/03/2020 to 20/05/2021.

**S5 Dataset. Daily COVID-19 Case Residuals.** England and Wales

**S6 Dataset. Daily COVID-19 Death Residuals.** England

**Fig S1. Daily histograms of LTLA COVID-19 cases per capita.** Black line represents the generalised logistic distribution and the red dashed line represents the normal distribution.

**Fig S2. Daily LTLA Density scaling behaviour of COVID-19 cases (i.e. log(Case Density vs. log(Population Density)).** The blue dots are the empirical values. A red line represents the single exponent power-law fit.

**Fig S3. Daily LTLA Density scaling plots of COVID-19 death (i.e. log(Death Density vs. log(Population Density)).** The blue dots are the empirical values (England). A red line represents the single exponent power-law fit.

**Fig S4. Daily geoplots of LTLA COVID-19 case residuals.** Regions that are red are above expectation and blue is below. The darker the shade the further from the scaling law.

**Fig S5. Daily geoplots of LTLA COVID-19 death residuals.** Regions that are red are above expectation and blue is below. The darker the shade the further from the scaling law.

**Fig S6. Spearman's rank correlation coefficient.** Rank correlation of COVID-19 case residuals between all pairs of dates between 01/03/2020 and 20/05/2021. Red and blue colours refer to correlation values close to 1 and -1 respectively. The darker shade of colour is associated to a higher similarity between the two pairs.

**Fig S7. Daily histograms of LTLA COVID-19 case residuals.** Black line represents the generalised logistic distribution and the red dashed line represents the normal distribution.

**Fig S8. Daily histograms of LTLA COVID-19 case deaths.** Black line represents the generalised logistic distribution and the red dashed line represents the normal distribution.